\documentclass{ws-procs961x669}            % default, citations in superscript
\begin{document}
\title{Scaling Solutions of Wiggly Cosmic Strings}

\author{A. R. R. Almeida$^*$}
\address{Centro de Astrof\'{\i}sica da Universidade do Porto, and\\
Instituto de Astrof\'{\i}sica e Ci\^encias do Espa\c co, Universidade do Porto,\\ Rua das Estrelas, 4150-762 Porto, Portugal, and\\
Faculdade de Ciências da Universidade do Porto\\
 Rua do Campo Alegre 687, 4169-007 Porto, Portugal\\
$^*$E-mail: Ana.Almeida@astro.up.pt}

\author{C. J. A. P. Martins$^+$}
\address{Centro de Astrof\'{\i}sica da Universidade do Porto, and\\
Instituto de Astrof\'{\i}sica e Ci\^encias do Espa\c co, Universidade do Porto,\\ Rua das Estrelas, 4150-762 Porto, Portugal\\
$^+$E-mail: Carlos.Martins@astro.up.pt}

\begin{abstract}
Cosmic string networks form during cosmological phase transitions as a consequence of the Kibble mechanism. The evolution of the simplest networks is accurately described by the canonical Velocity Dependent One-Scale (VOS) model. However, numerical simulations have demonstrated the existence of  significant quantities of short-wavelength propagation modes on the strings, known as wiggles, which motivated the recent development of a wiggly string extension of the VOS. Here we summarize recent progress in the physical interpretation of this model through a systematic study of the allowed asymptotic scaling solutions of the model. The modeling mainly relies on three mechanisms: the universe's expansion rate, energy transfer mechanisms (e.g., the production of loops and wiggles), and the choice of the scale in which wiggles are coarse-grained. We consider the various limits in which each mechanism dominates and compare the scaling solutions for each case, in order to gain insight into the role of each mechanism in the overall behavior of the network. Our results show that there are three scaling regimes for the wiggliness, consisting of the well-known Nambu-Goto solution, and non-trivial regimes where the amount of wiggliness can grow as the network evolves or, for specific expansion rates, become a constant. We also demonstrate that full scaling of the network is more likely in the matter era than in the radiation epoch, in agreement with numerical simulations.
\end{abstract}


\bodymatter

\section{Introduction}

Cosmic string networks arise in many theories of unification beyond the standard model. In the context of cosmology, they are predicted to have been formed in the early universe as a consequence of the Kibble mechanism \cite{kibble1976}. Due to the non-linearity and non-trivial interactions intrinsic to these networks, a complete quantitative description of their evolution, as well as of their observational consequences, proves to be a difficult task \cite{VSH}. 

An accurate description of cosmic strings is accomplished through two complementary approaches: numerical simulations (Nambu-Goto or Abelian-Higgs) and analytic models. The analytic modeling of string networks relies on the statistical physics of the network so as to derive a description of its thermodynamics. The analytic framework for the simplest Nambu-Goto string networks is the Velocity Dependent One-Scale (VOS) model \cite{martins1996,martins1996scale}. As the direct comparisons with Nambu-Goto and Abelian-Higgs numerical simulations \cite{moore2001, martins2006fractal, Correia1, Correia2} demonstrate, the VOS is able to capture the key large-scales properties of the network extremely well. The VOS is built upon the one-scale assumption embodied in the original model developed by Kibble \cite{kibble}, which expresses the evolution of the network in terms of a single length scale. This length scale can be the inter-string distance, the correlation length $\xi$, or the average curvature radius $R$ (note these length scales are expected to be identical in this formalism). Various frameworks attempting to describe the dynamics of string defects have been developed, like the three-scale approach \cite{austin1993}, but it was only with the advent of the VOS that a model capable of making robust predictions in various cosmological epochs emerged. The key difference in the VOS framework lies in an additional macroscopic quantity, a root-mean-square (RMS) velocity, whose evolution is described by an extension of the Newton's Second Law. The equations of the standard Nambu-Goto VOS model are rigorously deduced from the Nambu-Goto effective action.

In spite of the successes of the VOS, the model fails to give an accurate description of realistic networks of strings, whose worldsheets are expected to have additional degrees of freedom. In particular, the aforementioned numerical simulations of cosmic strings in expanding universes have demonstrated the existence of great amounts of short-wavelength propagation modes on the strings, known as wiggles, on scales orders of magnitude below the correlation length. The presence of small-scale structure is simply a byproduct of the energy loss phenomena of the network, and thus, if one aims to have an accurate analytic framework of string networks, it is imperative that these small-scale wiggles are accounted for. This motivated the development of a wiggly extension of the VOS that describes the evolution of both large-scale and small-scale properties of string networks \cite{PAP1,PAP2}. 

Here we continue the work recently reported in Ref. \citenum{almeida2021scaling} studying the properties of wiggly cosmic strings, through an improvement of its physical interpretation and modeling. We carry out a systematic study of the asymptotic scaling solutions of the wiggly generalization of the VOS, so as to explore the allowed scaling regimes, other than the trivial Nambu-Goto solution, where wiggliness grows, disappears, or evolves towards a constant value. One of the open questions in the wiggly model is whether the wiggliness reaches scaling, and if so, under what physical conditions. Numerical studies \cite{martins2006fractal} hint at scaling being reached at least in the matter-dominated era.

Another aim of this mathematical exploration is to determine the role that each mechanism has in the overall behavior of the network, as we know the nature of the scaling solutions is highly dependent on the dominant physical mechanism \cite{PAP2}. We consider the various limits in which each mechanism dominates and compare the scaling solutions for each case. Finally, this study also aims to pave the way for a future calibration of the wiggly model. Having determined the various possible scaling regimes, and particularly the behavior of the small-scale structure, we can plan future field theory and Nambu-Goto simulations that are optimized for this calibration and able to test the consistency relations of the solutions, as well as to make numerical predictions of the wiggliness on the strings.

This paper is organized as follows. Section 2 gives a brief overview of the wiggly extension of the VOS. In Section 3, we discuss the asymptotic scaling solutions without and with cosmological expansion, followed by a discussion of their physical implications. Finally we present some conclusions and an outlook in Section 4.

\section{Wiggly Strings Dynamics}

A wiggly extension of the VOS is accomplished by noting that, in direct contrast with Nambu-Goto strings, wiggly strings have their local string tension $T$ differ from the energy density in the string's local rest frame $U$. In particular, we can define a parameter $w$ that ranges from $0$ to $1$ (unity being the value of a Nambu-Goto string), such that 
\begin{equation}
\frac{T}{\mu_o}=w, \quad \frac{U}{\mu_o}=\frac{1}{w},
\label{wiggly1}
\end{equation}
where $\mu_o$ is the string mass per unit length. The total energy of a piece of string is given by
\begin{equation}
E=a \int \epsilon U d\sigma =\mu_o a\int\frac{\epsilon}{w}d\sigma,
\label{wiggly2}
\end{equation}
which can be decomposed into two contributions: one for the bare string segments $E_o$, 
\begin{equation}
E_o=\mu_o a \int\epsilon d\sigma,
\label{wiggly3}
\end{equation}
and another due to the small-scale wiggles $E_w$: 
\begin{equation}
E_w= \mu_o a\int\frac{1-w}{w}\epsilon d\sigma.
\label{wiggly4}
\end{equation}

Wiggly strings can be conceived as containing a mass current that renormalizes the bare energy per unit length \cite{martins2016defect}. We define the renormalization factor
\begin{equation}
\mu \equiv \frac{E}{E_o} \geq 1,
\label{wiggly5}    
\end{equation}
which is a measure of the energy due to the wiggles of the network, known as the renormalized string mass per unit length. $\mu$ is a scale-dependent quantity, that is, it depends on the renormalization scale $\ell$ at which the wiggliness is being measured. Each energy contribution yields a characteristic length scale
\begin{equation}
\rho\equiv\frac{\mu_o}{L^2}
\label{wiggly6} 
\end{equation}
\begin{equation}
\rho_o \equiv\frac{\mu_o}{\xi^2},
\label{wiggly7} 
\end{equation}
which are related by
\begin{equation}
\mu = \bigg( \frac{\xi}{L} \bigg)^2.
\label{wiggly8} 
\end{equation}
This procedure makes it clear that a quantitative description of small-scale structure implies that the one-scale assumption implicit to the VOS no longer holds. Instead, we have two distinct length scales $L$ and $\xi$, which will have different dynamics as it will become apparent in their evolution equations. Thus, an averaged model for wiggly cosmic string evolution implies three evolution equations, instead of two, with this additional equation accounting for the evolution of the network's small-scale structure. We note the assumption of $\xi \sim R$ still holds in the wiggly formalism, but it can be tested numerically \cite{martins2006fractal}. It should also be noted that while $\xi$ measures a physical length, $L$ is the length that a Nambu-Goto string with the same total energy would have, and thus should not be interpreted as a physical length.

The equations for the evolution of the network's dynamical quantities are obtained from averaging the microscopic string equations of motion, and so one needs to define the averaging procedure. There are two natural approaches to this. The first consists of averaging over the total energy, so as to give more weight to string segments with larger mass currents. For a generic quantity $Q$, this is defined as:
\begin{equation}
\langle Q\rangle=\frac{\int Q \frac{\epsilon}{w} d \sigma}{\int \frac{\epsilon}{w} d \sigma} 
\end{equation}
An alternative approach consists of averaging over the bare string energy
\begin{equation}
\langle Q\rangle_{0}=\frac{\int Q \epsilon d \sigma}{\int \epsilon d \sigma}
\end{equation}
The two procedures are related by: 
\begin{equation}
\langle Q\rangle=\frac{\langle Q U\rangle_{0}}{\langle U\rangle_{0}}=\frac{\langle Q / w\rangle_{0}}{\mu}.
\label{wiggly9.2}
\end{equation}
In the wiggly formalism \cite{PAP1, PAP2} the first averaging method is the one adopted (and is the one we adopt in what follows).

The various interactions by which the network loses energy still need to be discussed and included in the evolution equations. In particular, we need to define the phenomenological terms that account for the energy transfers between the bare string and the wiggles. While in the standard VOS model, long string intercommutings had no effect on the evolution of the network, in wiggly strings, they lead to the formation of kinks that increase the energy due to wiggles, and thus, need to be taken into consideration. The energy transfer from the bare string to the wiggles is modeled by:
\begin{equation}
\left(\frac{1}{\rho_o}\frac{d\rho_o}{dt}\right)_{\rm wiggles}= -cs(\mu)\frac{v}{\xi},
\label{wiggly10}
\end{equation}
in which $s$ approaches 0 in the Nambu-Goto limit. Besides accounting for the gain in small-scale structure, $s$ should also include the effects of kink decay by gravitational radiation.

Another mechanism by which the network loses energy is the production of loops. In the VOS model, this is encoded in a phenomenological parameter $c$, the loop chopping efficiency. The loop chopping parameters of the wiggly model are defined in analogy with the VOS definition, but now the energy transferred to the loops may come from the bare string or from the wiggles
\begin{equation}
\left(\frac{1}{\rho_o}\frac{d\rho_o}{dt}\right)_{\rm loops}=- c f_o(\mu)\frac{v}{\xi},
\label{wiggly11}
\end{equation}
\begin{equation}
\left(\frac{1}{\rho_w}\frac{d\rho_w}{dt}\right)_{\rm loops}=-c f_1(\mu) \frac{v}{\xi},
\label{wiggly12}
\end{equation}
such that 
\begin{equation}
\left(\frac{1}{\rho}\frac{d\rho}{dt}\right)_{\rm loops}=\left(\frac{1}{\rho}\frac{d\rho_0}{dt}\right)_{\rm loops}+\left(\frac{1}{\rho}\frac{d\rho_w}{dt}\right)_{\rm loops}.
\label{wiggly13}
\end{equation}
For simplicity, one can define an overall energy loss parameter
\begin{equation}
\left(\frac{1}{\rho}\frac{d\rho}{dt}\right)_{\rm loops} \equiv -c f(\mu) \frac{v}{\xi},
\label{wiggly14}
\end{equation}
which using equations (\ref{wiggly11}--\ref{wiggly13}) can be shown to yield:
\begin{equation}
f(\mu) =\frac{f_o(\mu)}{\mu}+ \left(1-\frac{1}{\mu}\right) f_1(\mu).
\label{wiggly15}
\end{equation}
Note that $f$ has an explicit dependence on $\mu$ so as to account for the fact that loop production is favored in regions of the long string network with more small-scale structure than average \cite{allen1990cosmic, bennett1990high, martins2006fractal}. In what follows we adopt the assumptions introduced in Ref. \citenum{PAP2}
\begin{equation}
\begin{aligned}
f_{o}(\mu) &=1 \\
f(\mu) &=1+\eta\left(1-\frac{1}{\sqrt{\mu}}\right) \\
s(\mu) &=D\left(1-\frac{1}{\mu^{2}}\right),
\label{wiggly16}
\end{aligned}
\end{equation}
where two additional phenomenological parameters $\eta$ and $D$ were introduced, which should be interpreted as probabilities for small-scale structure loss and gain, respectively.

Finally, we emphasize that the renormalized string mass per unit
length is defined at a renormalization scale $\ell$ that need
not be fixed but can be time dependent (e.g., the correlation length). In fact, the choice of scale is equivalent to altering what defines a wiggle, and consequently, must have a direct influence on the energy distribution between the bare string and the wiggles. One accounts for this by introducing the following scale-drift terms
\begin{equation}
\frac{1}{\mu} \frac{\partial \mu}{\partial \ell} \frac{d \ell}{d t} \sim \frac{d_{m}-1}{l} \frac{d \ell}{d t},
\label{wiggly17}
\end{equation}
\begin{equation}
\frac{\partial v^{2}}{\partial \ell} \frac{d \ell}{d t}=\frac{1-v^{2}}{1+\left\langle w^{2}\right\rangle} \frac{\partial\left\langle w^{2}\right\rangle}{\partial \ell} \frac{d \ell}{d t},
\label{wiggly21}
\end{equation}
where $d_m(\ell)$ is the multifractal dimension of a string segment at scale $\ell$. It should be emphasized that while (\ref{wiggly17}) is a mere geometric identity, (\ref{wiggly21}) is a consequence of ensuring energy conservation at all scales. We will also assume the following phenomenological relation 
\begin{equation}
d_m(\mu)=2-\frac{1}{\mu^b},
\label{wiggly20}
\end{equation}
where $b$ is a free parameter that imposes the transition limit between two regimes ($d_m=2$ for Brownian networks, $d_m=1$ for small scales). 

We are now ready to derive the system of equations that compose the wiggly model \cite{PAP1,PAP2}
\begin{equation}
2\frac{dL}{dt}=HL\left[3+v^2-\frac{(1-v^2)}{\mu^2}\right]+\frac{cfv}{\sqrt{\mu}}
\label{wiggly22}
\end{equation}
\begin{equation}
2\frac{d\xi}{dt}=H\xi\left[2+\left(1+\frac{1}{\mu^{2}}\right)v^{2}\right]+v\left[k\left(1-\frac{1}{\mu^{2}}\right)+c\left(f_{o}+s\right)\right]
+\left[d_{m}\left(\ell \right)-1\right]\frac{\xi}{\ell}\frac{d \ell}{dt}
\label{wiggly23}
\end{equation}
\begin{equation}
\frac{dv}{dt}=\left(1-v^{2}\right)\left[\frac{k}{\xi\mu^{2}}-Hv\left(1+\frac{1}{\mu^{2}}\right)-\frac{1}{1+\mu^{2}}\frac{\left[d_{m}\left( \ell \right)-1\right]}{v \ell}\frac{d \ell}{dt}\right]
\label{wiggly24}
\end{equation}
\begin{equation}
\frac{1}{\mu}\frac{d\mu}{dt}=\frac{v}{\xi}\left[k\left(1-\frac{1}{\mu^{2}}\right)-c\left(f-f_{o}-s\right)\right]-H\left(1-\frac{1}{\mu^{2}}\right)
+\frac{\left[d_{m}\left(\ell \right)-1\right]}{\ell}\frac{d \ell}{dt},
\label{wiggly25}
\end{equation}
where $H \equiv {\Dot{a}}/{a}$ is the Hubble parameter (which we will assume to be of the form $a\propto t^\lambda$) and $k$ the momentum parameter already introduced in the VOS formalism. $k$ essentially quantifies the local curvature of strings. It should be noted that equations (\ref{wiggly22}), (\ref{wiggly23}) and (\ref{wiggly25}) are not independent, being related by Eq. (\ref{wiggly8}). 

\section{Scaling Solutions}

We carry out a mathematical exploration of the landscape of scaling solutions of the wiggly model, so as to characterize the behavior of the wiggliness of the network. One important yet unanswered question in the wiggly model is whether the wiggliness is able to scale, i.e, evolve towards a constant value, and if so, what are the underlying physical conditions of the network that lead to this behavior. In spite of current numerical simulations not giving a definitive answer to this question \cite{martins2006fractal}, they hint at scaling being reached in the matter-dominated era ($\lambda = 2/3$), with the behavior in the radiation era being more complex. This may be indicative of an absence of a scaling solution or of a slower scaling regime (since there is less Hubble damping).

As shown by the system of equations (\ref{wiggly22})- (\ref{wiggly25}), the evolution of the network is driven by three main mechanisms: the universe’s expansion, energy loss by intercommutation, and the choice of the scale in which wiggles are coarse-grained. In what follows, we study the effects of each of them on the possible asymptotic scaling solutions by considering the case where each mechanism dominates and by doing a subsequent comparison of the obtained scaling solutions in each case. This allows us to gain insight into the role that each mechanism has on the evolution of the network. 

From a physical point of view, we expect three types of scaling solution regimes. Firstly, we evidently expect the Nambu-Goto solution $\mu = 1$, as the wiggly model reduces to the VOS in the appropriate limit. Secondly, we also expect a regime where the wiggliness scales, $\mu = m_o$. Finally, a regime where small-scale structure grows $\mu \propto t^{\gamma}$ as the network evolves is also expected. In general we will assume
\begin{equation}
\begin{aligned}
L &=L_o t^{\alpha} \\
v &=v_o t^{\beta} \\
\mu &=m_o t^{\gamma},
\label{scal1}
\end{aligned}
\end{equation}
or alternatively, using Eq. (\ref{wiggly8}):
\begin{equation}
\xi = L_o m_o^{1/2}  t^{\alpha +\gamma/2}.
\label{scal2}
\end{equation}
It should also be noted that we will discard ultra-relativistic $v = 1$ solutions in our analysis, as well as other branches of solutions that are mathematically allowed but have no physical meaning. This includes solutions in Minkowski space $H =0$ (expanding universes $H \neq 0$) that hold for non-null values of the momentum parameter $k \neq 0$ (null values of momentum parameter $k=0$). The reason behind this is we expect the VOS to hold for $k=0$ in Minkowski space \cite{martins2004unified,martins2016defect,martins2006fractal,Correia1}, and in an analogous manner, we expect $k \neq 0$ when $H \neq 0$ \cite{martins2006fractal}.

\subsection{Scaling Solutions Without Expansion}

First, we consider the allowed scaling solutions in Minkowski space. We note that linear scaling has been observed in numerical simulations in Minkowski space \cite{martins2006fractal,sakellariadou1990cosmic}. 
We also note that a more detailed analysis of some of these solutions has already been published in Ref. \citenum{almeida2021scaling}. 

\subsubsection{No dynamical mechanism}

We start by considering the simplest case where no dynamical mechanisms act on the network. In this case we find a trivial solution that depicts a network where all four dynamical quantities remain constant but undetermined
\begin{equation}
\begin{aligned}
L &=L_o \\
v &=v_o \\
\mu &=m_o \\
\xi &=\sqrt{m_o} L_o.
\label{scal3.1} 
\end{aligned}
\end{equation}
The values of these quantities could be determined by measurement in numerical simulations. The physical interpretation of this solution is clear: the network is in equilibrium due to the absence of energy loss mechanisms, and thus its dynamical quantities have no time evolution.

\subsubsection{With running averaging scale}

In this case we assume the fractal dimension introduced in Eq. (\ref{wiggly20}) with the specific choice $b = 2$, as numerical simulations suggest this value is reasonable \cite{martins2006fractal}. We also assume a generic power-law shaped averaging scale 
\begin{equation}
l = l_o t^{\delta}, \quad 0 < \delta \leq 1.
\label{scal4.1}
\end{equation}
In the absence of energy loss mechanisms, we expect that the characteristic length scale of the network remains constant. We find two distinct scaling solutions, the first being the Nambu-Goto limit
\begin{equation}
\begin{aligned}
L &=L_o \\
v &=v_o \\
\mu &= 1 \\
\xi &= L_o,
\label{scal4.2}
\end{aligned}
\end{equation}
which is analogous to that of  Eq.(\ref{scal3.1}), with the difference that $\mu$ is no longer arbitrary but restricted to the Nambu-Goto value. This solution gives a clear depiction of how, when there is no small-scale structure, a variation in the coarse-graining scale makes no difference. In other words, this solution exists for any value of $\delta$ and has no explicit dependence on its choice.

The second solution yields growing wiggliness
\begin{equation}
\begin{aligned}
L &=L_o \\
v &=v_o t^{-\delta} \\
\mu &=  m_o t^{\delta} \\
\xi &= \sqrt{m_o} L_o t^{\delta/2},
\label{scal4.3}
\end{aligned}
\end{equation}
subject to the following constraint:
\begin{equation}
v_o^2 m_o^2 =1.
\label{scal4.4}
\end{equation}
We note that in the fixed scale limit $\delta \longrightarrow 0$ we recover the scaling exponents of Eq. (\ref{scal4.1}). In addition, note that choosing our renormalization scale to be the correlation length $l \propto \xi$, yields $\delta= {\delta}/{2}=0$, implying $\gamma=0$, or in other words, the only possible solution in this case is the Nambu-Goto one. The physical interpretation of this solution is clear. The presence of small-scale structure made this solution explicitly dependent on $\delta$. In particular, we are always able to find small-scale structure for all choices of scale, and a variation in the scale simply modifies the way energy migrates from the bare string to the wiggles, with the total energy of the network staying constant. Moreover, the growth of the wiggliness is compensated by a decrease in the velocity, which suggests that not only is $\mu$ scale-dependent, but $v$ also depends on $\delta$. This agrees with the mesoscopic velocity interpretation of the wiggly model, instead of the traditional interpretation of a microscopic RMS one embodied in the VOS \cite{PAP1,PAP2}.

\subsubsection{With energy losses}

We now consider the case where the network is losing energy due the various energy loss mechanisms inherent to the network, as previously described. In particular, we assume the energy loss terms are modeled after Eq. (\ref{wiggly16}). It is clear that the presence of energy losses must lead to a different behavior for the network's characteristic length scales $L$ and $\xi$, as they are now forced to scale in time. We find two solution regimes. The first solution implies non-trivial wiggliness:
\begin{equation}
\begin{aligned}
L &=L_o t \\
v &=v_o \\
\mu &=  m_o \\
\xi &= \sqrt{m_o} L_o t,
\label{scal5.1}
\end{aligned}
\end{equation}
such that
\begin{equation}
L_o =\frac{cv_o}{2} \frac{(1+\eta)m_o^{1/2}-\eta}{m_o}
\label{scal5.2}
\end{equation}
\begin{equation}
D\left(1-\frac{1}{m_{o}^{2}}\right)=\eta\left(1-\frac{1}{m_o^{1/2}}\right), \quad \eta \geq D.
\label{scal5.3}
\end{equation}
This allows for a wide range of values of the wiggliness $m_o$, depending on the values of the small-scale structure parameters $D$ and $\eta$. In particular, one retrieves the Nambu-Goto case $m_o=1$  under the condition $D=\eta$. For realistic values of $D$ and $\eta$, Eq. (\ref{scal5.3}) has two real solutions and a pair of complex conjugate solutions. Having in mind that realistic values of the wiggliness imply $m_o \geq 1$, it follows that for values of the parameters such that ${D}/{\eta} < {1}/{4}$ the only physically meaningful solution consists of $m_o=1$, while in the range ${D}/{\eta} \in \left] {1}/{4}, 1 \right[$, we find solutions with  $m_o>1$. From a physical point of view, this should be interpreted as: if the amount of wiggliness generated by the network is a small fraction of the amount that is lost, as in the case of ${D}/{\eta} < {1}/{4}$, then no wiggliness can asymptotically survive on the string network, resulting in the Nambu-Goto solution. On the other hand, if this ratio takes up large enough values, then the small-scale structure of the network is able to survive.

In the second regime, we find a solution where small-scale structure builds-up in the network:
\begin{equation}
\begin{aligned}
L &=L_o t^{\alpha} \\
v &=v_o \\
\mu &=  m_o t^{2-2\alpha} \\
\xi &= \sqrt{m_o} L_o t
\label{scal5.4}
\end{aligned}
\end{equation}
with the following constraints and conditions
\begin{equation}
L_o= \frac{cv_o}{2\alpha}\frac{1+\eta}{m_o^{1/2}}
\label{scal5.5}
\end{equation}
\begin{equation}
\alpha=\frac{1+\eta}{1+D} < 1
\label{scal5.6}
\end{equation}
\begin{equation}
\gamma= 2 \frac{D-\eta}{1+D}, \quad D>\eta,
\label{scal5.7}
\end{equation}
which yields Eq. (\ref{scal5.1}) in the limit where the two small-scale structure parameters balance each other $D \longrightarrow \eta$. 

The set of solutions (\ref{scal5.1}) (\ref{scal5.4}) allows us to infer the exact physical conditions that dictate the nature of the scaling regime. As seen in Eq. (\ref{scal5.1}), when the amount of small-scale structure removed (e.g., loop production) is greater than its gain (e.g., kink formation upon intercommutation), $\eta > D$, the wiggliness of the network remains constant. On the other hand, as illustrated by Eq. (\ref{scal5.4}), when there is more generation of small-scale structure on the network than loss, $D > \eta$, we find that the long string wiggliness must grow. Moreover, another interesting difference between these two solutions lies in the behavior of $\mu$ and $L$ (or equivalently, the total energy density of the network), with $\xi$ scaling linearly ($\xi \propto t$) in both solutions. The fact that $\xi$ maintains its linear scaling independently of the scaling of $L$ just further emphasizes the two scale feature of the wiggly VOS extension.

\subsubsection{With energy losses and a running averaging scale}

If one allows for both energy losses and a varying renormalization scale $\delta \neq 0$, one finds two possible solutions. The first yields constant wiggliness such that:
\begin{equation}
\begin{aligned}
L &= \frac{c(1+\eta)v_o}{2} t \\
v &= v_o  \\
\mu &=  1\\
\xi &= L,
\label{scal7.1}
\end{aligned}
\end{equation}
which is analogous to that of Eq.(\ref{scal5.1}) with the caveat that $m_o$ is no longer arbitrary but restricted to the Nambu-Goto case. This solution depicts, yet again, that if the network is devoid of small-scale structure a change in the coarse-graining scale makes no difference.

In the second solution, we find that the small-scale structure of the network grows as the network evolves
\begin{equation}
\begin{aligned}
L &= L_o t^{\alpha} \\
v &= v_o t^{-\gamma}  \\
\mu &=  m_o t^{\gamma}\\
\xi &= \sqrt{m_o} L_o t^{\frac{1}{3}+\frac{2}{3}\alpha},
\label{scal7.3}
\end{aligned}
\end{equation}
subject to
\begin{equation}
L_o =\frac{c(1+\eta)v_o}{2\alpha m_o^{1/2}}
\label{scal7.4}
\end{equation}
\begin{equation}
\alpha = \frac{(\frac{2}{3}-\delta)(1+\eta)}{\frac{2}{3}(1+\eta)+2(D-\eta)} < 1, \quad  D>\eta 
\label{scal7.5}
\end{equation}
\begin{equation}
\gamma =  \frac{2}{3}-\frac{2}{3}\alpha = \frac{2(D-\eta)+\delta(1+\eta)}{3(D-\eta)+(1+\eta)}
\label{scal7.6}
\end{equation}
\begin{equation}
\delta =\gamma v_o^{2} m_o^{2} \in \left]0, \frac{2}{3} \right[
\label{scal7.7}
\end{equation}
Note that if we consider the fixed scale limit $\delta \longrightarrow 0$, we recover Eq. (\ref{scal5.4}) $\gamma \longrightarrow 0$, $\alpha \longrightarrow 1$. We can also take the limit of no energy losses $D, \eta \longrightarrow 0$ on (\ref{scal7.3}) where we partially recover the previous solution of Eq. (\ref{scal4.3}), namely the dynamics for the velocity and wiggliness $ \gamma = - \beta = \delta$, but with different behavior of both length scales.  Unlike the other solutions so far, $\delta$ is restricted because Eq. (\ref{scal7.7}) must always hold. Specifically, the averaging scale is now bounded from above, which can be attributed to the energy losses of the network. In other words, if the scale does not evolve slowly enough, we are unable to find small-scale structure. Furthermore, we note that the total energy of the network, more specifically the exponent $\alpha$, explicitly depends on $\delta$, hinting at an apparent scale dependence of the total energy of the network. However, one would expect this quantity to be scale-invariant, that is, the way the network loses energy should not depend on the choice of averaging scale choice. Having said this, it is clear that the physics behind this behavior needs to be explored in future studies.

\subsection{Scaling Solutions With Expansion}

We are now concerned with solutions in expanding universes, where we expect $k \neq 0$ \cite{martins2006fractal}.

\subsubsection{No dynamical mechanisms}

We start with the case of no energy losses and a fixed scale. In this case, we find the three expansion regimes. Firstly, we find the canonical Nambu-Goto VOS solution
\begin{equation}
\begin{aligned}
L &=\bigg( \frac{k^2}{4\lambda (1-\lambda)} \bigg) ^{1/2} t \\
v &= \bigg( \frac{1-\lambda}{\lambda} \bigg) ^{1/2} \\
\mu &=  1 \\
\xi &= \bigg( \frac{k^2}{4\lambda (1-\lambda)} \bigg) ^{1/2} t,
\label{scal6.2}
\end{aligned}
\end{equation}
The range of expansion rates of this solution depends on the physical interpretation of the velocity. Interpreted as a microscopic velocity, we require $ v_o^2 < 1$  and we find that the radiation epoch is the limiting case $ \lambda > {1}/{2}$. This implies that scaling can be achieved in the matter-dominated era, but not in the radiation epoch, since in the latter the network does not lose enough energy by Hubble damping. Alternatively, one can also interpret this velocity as an average one, with now the physical constraint being  $ v_o^2 \leq {1}/{2}$ (corresponding to the average velocity of loops in Minkowski space), leading to the matter era becoming the limiting case $\lambda \geq {2}/{3}$.

Secondly, we find a solution of non-trivial constant wiggliness, only viable in the matter era
\begin{equation}
\begin{aligned}
L &= L_o t \\
v &= v_o \\
\mu &=  m_o \\
\xi &= \sqrt{m_o} L\\
\lambda &= \frac{2}{3}
\label{scal6.5}
\end{aligned}
\end{equation}
\begin{equation}
L_o= \frac{3k}{2(m_o^3+m_o)^{1/2}}, \quad v_o= \frac{1}{(1+m_o^2)^{1/2}},
\label{scal6.6}
\end{equation}
which in principle holds for any value of the wiggliness $m_o$. We note that by choosing $m_o=1$, we retrieve the previous solution (\ref{scal6.2}). This solution also demonstrates that scaling is more easily reached in the matter era than in the radiation era, in agreement with Nambu-Goto numerical simulations \cite{martins2006fractal}. In addition, the behavior depicted by this solution is analogous to the one previously identified for chiral superconducting strings \cite{oliv}.

Thirdly, a regime of growing wiggliness is also allowed, which implies decreasing velocities $\beta < 0$, and $\alpha ={3}/{2}\lambda$, as well as the consistency relations  ${3}/{2} \lambda + {\gamma}/{2}-\beta=1$, and $\beta+\gamma \geq 0$. This narrows the range of allowed expansion rates to slower rates than the matter era one $\lambda < {2}/{3}$. There are two solution regimes, depending on the value of the expansion rate. For very slow expansion rates we have
\begin{equation}
\begin{aligned}
L &= L_o t^{ 3/2\lambda} \\
v &= v_o t^{-\lambda} \\
\mu &=  m_o t^{2 -5\lambda}\\
\xi &= \frac{v_o k}{2-4\lambda} t^{1-\lambda},
\label{scal6.7}
\end{aligned}
\end{equation}
subject to the following constraints:
\begin{equation}
m_o = \bigg( \frac{v_o k}{L_o(2-4\lambda)} \bigg)^{2}
\label{scal6.8} 
\end{equation}
\begin{equation}
\lambda \leq \frac{1}{3}.
\label{scal6.9}    
\end{equation}
Conversely, in the intermediate expansion rate regime, we find
\begin{equation}
\begin{aligned}
L &= L_o t^{ 3/2\lambda} \\
v &= v_o t^{-\gamma} \\
\mu &= m_o t^{\gamma}\\
\xi &= \sqrt{m_o} L_o t^{\lambda+\frac{1}{3}}\\
\gamma &= \frac{2}{3} -\lambda,
\label{scal6.12}
\end{aligned}
\end{equation}
with the following constraints:
\begin{equation}
v_o = \frac{2}{3} \bigg( \frac{L_o}{(\frac{4}{3}\lambda - \frac{4}{9})^{1/4} k} \bigg)^{2/3}, \quad m_o= \bigg( \frac{ k }{(\frac{4}{3}\lambda - \frac{4}{9})^{1/2} L_o } \bigg)^{2/3}
\label{scal6.13}
\end{equation}
\begin{equation}
\lambda \in \bigg] \frac{1}{3}, \frac{2}{3} \bigg[.
\label{scal6.14}
\end{equation}
We note that in the limit $\lambda \longrightarrow {1}/{3}$ this solution matches the scaling solution obtained in the slow regime. In addition, we also recover the full scaling solution (\ref{scal6.5}) in the limit $\lambda \longrightarrow {2}/{3}$, apart from differences in the coefficients for the velocity. 

It is instructive to consider the particular case of the radiation epoch ($\lambda={1}/{2}$), where we find
\begin{equation}
L \propto t^{3/4}, \quad v \propto t^{-1/6} \quad \mu \propto t^{1/6} \quad \xi \propto \sqrt{m_o} L_o t^{5/6},
\label{scal6.17}
\end{equation}
These relations provide an opportunity for testing with Nambu-Goto simulations where energy loss terms can be switched off on demand.

\subsubsection{With a running averaging scale}

We consider the averaging scale previously introduced in Eq. (\ref{scal4.1}). In this case we still find the Nambu-Goto solution
\begin{equation}
\begin{aligned}
L &= L_o t \\
v &= v_o \\
\mu &=  1\\
\xi &= \sqrt{m_o} L_o t\\
\lambda &> \frac{1}{2},
\label{scal8.1}
\end{aligned}
\end{equation}
\begin{equation}
L_o = \bigg( \frac{k^2}{4\lambda (1-\lambda)} \bigg) ^{1/2}, \quad v= \bigg( \frac{1-\lambda}{\lambda} \bigg) ^{1/2}.
\label{scal8.2}
\end{equation}
We observe that the introduction of a running averaging scale neither affected the scaling behavior of the previous obtained solution (\ref{scal6.2}) nor did it impose any restriction on $\delta$.

We also find a full scaling solution
\begin{equation}
\begin{aligned}
L &= L_o t \\
v &= v_o \\
\mu &=  m_o\\
\xi &= \sqrt{m_o} L_o t\\
\lambda &< \frac{2}{3}
\label{scal8.3}
\end{aligned}
\end{equation}
\begin{equation}
v_o = \bigg( \frac{1+m_o^2(2 \lambda^{-1}-3)}{1+m_o^2} \bigg)^{1/2} 
\label{scal8.4}
\end{equation}
\begin{equation}
\delta=(1-\frac{3}{2} \lambda) m_o^2 (1+m_o^2).
\label{scal8.5}
\end{equation}
This solution demonstrates how the inclusion of a renormalization scale decreased the value of the expansion rate for which full scaling can occur: this is now achieved for expansion rates lower than the matter era one. 

Lastly, we find two possible solutions with growing wiggliness, which entail the same dynamics for the characteristic length scale and velocity as found for the expansion case, namely $\alpha ={3}\lambda/2$, $\beta < 0$, ${3}/{2} \lambda + {\gamma}/{2}-\beta=1$, and $\beta+\gamma \geq 0$. For very slow expansion rates we have
\begin{equation}
\begin{aligned}
L &= L_o t^{3/2\lambda} \\
v &= v_o t^{-\lambda} \\
\mu &=  m_o t^{2-5\lambda}\\
\xi &= \sqrt{m_o} L_o t^{1-\lambda}\\
\lambda &\leq \frac{1}{3},
\label{scal8.8}
\end{aligned}
\end{equation}
where the scaling coefficients are related by:
\begin{equation}
m_o = \bigg( \frac{k v_o}{L_o(2-4\lambda+\delta)}\bigg)^{2},
\label{scal8.9}    
\end{equation}
while in the intermediate expansion regime we determine:
\begin{equation}
\begin{aligned}
L &= L_o t^{3/2\lambda} \\
v &= v_o t^{-\gamma} \\
\mu &=  m_o t^{\gamma}\\
\xi &= \sqrt{m_o} L_o t^{1/3+\lambda}\\
\frac{1}{3} & <\lambda<\frac{2}{3},
\label{scal8.12}
\end{aligned}
\end{equation}
such that
\begin{equation}
v_o= \bigg(\frac{1}{(3\lambda-1)m_o^2} \bigg)^{1/2},  \quad m_o = \bigg( \frac{k v_o}{L_o( \frac{2}{3}+ \delta)} \bigg)^2 
\label{scal8.13}  
\end{equation}
\begin{equation}
\gamma= \frac{2}{3} - \lambda.
\label{scal8.14}    
\end{equation}
Note that the two regimes match for an expansion rate $\lambda = \frac{1}{3}$.

\subsubsection{With energy losses}

We finally consider the most realistic case, where the network is losing energy in the context of an expanding universe. Firstly, we recover the canonical VOS Nambu-Goto solution
\begin{equation}
\begin{aligned}
L &= L_o t \\
v &= v_o \\
\mu &=  1\\
\xi &=  L_o t,
\label{scal9.1}
\end{aligned}
\end{equation}
\begin{equation}
L_o= \bigg( \frac{k(k+c)}{4\lambda(1-\lambda)} \bigg)^{1/2}, \quad  v= \bigg( \frac{k(1-\lambda)}{\lambda(k+c)} \bigg)^{1/2}.
\label{scal9.2}
\end{equation}
We note that Eq. (\ref{scal6.2}) is a particular case of Eq. (\ref{scal9.1}) for which $c=0$, which can be physically interpreted as the network's intrinsic energy losses being unnecessary for the network to reach linear scaling because Hubble damping is enough to ensure this.

Secondly, there is also a non-trivial wiggliness solution 
\begin{equation}
\begin{aligned}
L &= L_o t \\
v &= v_o \\
\mu &=  m_o\\
\xi &=  \frac{k}{\lambda v_o(1+m_o^{2})} t,
\label{scal9.3}
\end{aligned}
\end{equation}
subject to the following constraints:
\begin{equation}
L_o^2 =  \frac{k}{\lambda} \frac{k+c(1+\eta)m_o^2- c \eta m_o^{3 / 2}}{ m_o(1+m_o^2)(\lambda+(2-3 \lambda) m_o^2)} 
\label{scal9.4}
\end{equation}
\begin{equation}
v_o^2 = \frac{k}{\lambda}\frac{\lambda+(2-3 \lambda) m_o^{2}}{(1+m_o^{2})(k+c(1+\eta)m_o^2- c \eta m_o^{3/2} )}
\label{scal9.5}
\end{equation}
\begin{equation}
\bigg( \lambda+m_o^{2}(2-3 \lambda)\bigg) \bigg( (k+c D)(1-\frac{1}{m_o^{2}})-c \eta (1-\frac{1}{m_o^{1/2}} ) \bigg) =\lambda \bigg(1-\frac{1}{m_o^{2}} \bigg) \bigg(k+ (1+\eta-\eta m_o^{-1 / 2}) m_o^{2} c \bigg) .
\label{scal9.6}
\end{equation}
We note that the previous solution (\ref{scal6.5}) is recovered by direct substitution of $\lambda ={2}/{3}$, and $c=0$ on this solution. So as to understand more about its physical implications, we consider the case study of large amounts of wiggliness. Before proceeding to this analysis, we define new parameters
\begin{equation}
\begin{aligned}
k_{eff} &\equiv k +c(D-\eta) \\
c_{eff} &\equiv c(1+\eta).
\label{scal9.11} 
\end{aligned}
\end{equation}
This is not only convenient, but also makes sense from a physical point of view: it is known that the presence of small-scale structure on the long strings modifies its typical curvature, as well as enhances energy losses \cite{PAP2}. The large wiggliness limit $m_o \gg 1$
yields:
\begin{equation}
\begin{aligned}
L_o &=\bigg( \frac{k c_{eff}}{\lambda m_o^3 (2-3 \lambda)} \bigg)^{1/2} \\
v_o &=\bigg( \frac{k (2-3 \lambda)}{c\lambda m_o^2 (1+\eta)} \bigg)^{1/2} \\
\lambda &= \frac{2 k_{eff}}{c_{eff}+ 3k_{eff}},
\end{aligned}
\label{scal9.13} 
\end{equation}
This solution clearly illustrates that when there are great amounts of wiggliness on the network, the overall energy density increases while the velocity decreases, as expected. In addition, it is interesting to note that in the radiation era, the effective parameters balance one another
\begin{equation}
k_{eff} = c_{eff}. 
\label{scal9.14}       
\end{equation}
This relation constitutes an opportunity for testing with Nambu-Goto numerical simulations.

Thirdly, there also exists a growing wiggliness regime. This has the same implications as found for the expanding case, namely a decreasing velocity $\beta < 0$, although $\alpha={3}\lambda/2$ no longer holds, and instead we have $\alpha> {3}\lambda/2$. Physically, this makes sense as the energy loss terms enable faster energy loss. We again distinguish two regimes according to their expansion rates. For slow expansion rates,
\begin{equation}
\begin{aligned}
L &= L_o t^{\alpha} \\
v &= v_o t^{-\lambda}\\
\mu &=  m_o t^{\gamma}\\
\xi &=  \sqrt{m_o} L_o t^{1-\lambda},
\label{scal9.15}
\end{aligned}
\end{equation}
\begin{equation}
\alpha = \frac{1}{2} \frac{2 c_{eff} +\lambda(3k_{eff}-c_{eff})}{k_{eff}+c_{eff}}
\label{scal9.16}  
\end{equation}
\begin{equation}
\gamma = 2(1-\lambda -\alpha) = \frac{2k_{eff}-\lambda(c_{eff}+5k_{eff} )}{k_{eff}+c_{eff}}
\label{scal9.17}  
\end{equation}
\begin{equation}
\gamma \in ]0, 2-5 \lambda[  
\label{scal9.18}  
\end{equation}
\begin{equation}
\lambda \leq \frac{k_{eff}}{3 k_{eff}+c_{eff}}.
\label{scal9.19}  
\end{equation}
It is worthy of note that one recovers the previous slow expansion solution (\ref{scal6.7}) in the limit $c \longrightarrow 0$. 

As far as the intermediate expansion rate regime is concerned, we find
\begin{equation}
\begin{aligned}
L &= L_o t^{\alpha} \\
v &= v_o t^{-\gamma}\\
\mu &=  m_o t^{\gamma}\\
\xi &=  \sqrt{m_o} L_o t^{\frac{1}{3}+\frac{2}{3}\alpha}
\label{scal9.20}
\end{aligned}
\end{equation}
\begin{equation}
L_o^2 = \frac{k c_{eff}}{(2\alpha-3\lambda) (\lambda + \frac{2}{3}\alpha - \frac{2}{3}) m_o^{3} }
\label{scal9.21}
\end{equation}
\begin{equation}
v_o^2 =  \frac{k(2\alpha-3\lambda)}{c_{eff} (\lambda + \frac{2}{3}\alpha - \frac{2}{3}) m_o^{2}}  
\label{scal9.22}
\end{equation}
\begin{equation}
\alpha = \frac{\frac{2}{3}c_{eff}+\lambda(3k_{eff}+c_{eff})}{\frac{2}{3}c_{eff}+2k_{eff}}
\label{scal9.23}
\end{equation}
\begin{equation}
\gamma = \frac{2}{3}-\frac{2}{3}\alpha=  \frac{2k_{eff} -\lambda(3k_{eff}+c_{eff})}{c_{eff}+3k_{eff}}
\label{scal9.24}
\end{equation}
\begin{equation}
\lambda \in \bigg] \frac{k_{eff}}{3 k_{eff}+c_{eff}}, \frac{2 k_{eff}}{3 k_{eff}+c_{eff}} \bigg[,
\label{scal9.25}
\end{equation}
where one is also able to retrieve the previous solution Eq. (\ref{scal6.12}) in the limit $c \longrightarrow 0$. It should be emphasized that the domain of the expansion rate is now broader, when in comparison to Eq. (\ref{scal6.12}), which is due to the two energy loss mechanisms acting on the network (not just Hubble damping). 

Note that the two regimes match for the expansion rate
\begin{equation}
\lambda = \frac{k_{eff}}{3 k_{eff} + c_{eff}},
\label{scal9.26}
\end{equation}
which is half of the expansion rate of Eq. (\ref{scal9.13}). We also note that in the no energy losses limit $c \longrightarrow 0$, the expansion rate approaches $\lambda = {1}/{3}$.

Finally, it is also instructive to consider a particular case study where the two effective parameter balance each other $ k_{eff} = c_{eff} $ which yields in the slow regime
\begin{equation}
\alpha_{slow} =  \frac{1+\lambda}{2}, \quad \gamma_{slow}  = 1 - 3\lambda, \quad \lambda \leq \frac{1}{4},
\end{equation}
while in the intermediate regime we find:
\begin{equation}
\alpha_{int} = \frac{1}{4}+ \frac{3}{2}\lambda, \quad 
\gamma_{int}  = \frac{1}{2} - \lambda, \quad \lambda \in \bigg]\frac{1}{4},\frac{1}{2} \bigg[.
\label{scal9.29}    
\end{equation}
This solution leads to full scaling in the radiation epoch and growing wiggliness regimes for slower expansion rates than the radiation one. The transition between slow and intermediate regimes takes place at $\lambda = {1}/{4}$.

\section{Conclusions and Further Work}

We have improved the physical interpretation of the wiggly generalization of the VOS model \cite{PAP1,PAP2}, building upon a mathematical exploration of the landscape of possible scaling solutions. Not only did we successfully characterize the behavior of the wiggliness in various regimes, but we also have managed to determine the effect that each dynamical mechanism has on the scaling solutions.

In Minkowski space, we find that the network is in a trivial equilibrium solution in the absence of energy loss mechanisms. Allowing the network to lose energy, we find that the network's correlation length always maintains its linear scaling, while the behavior of the characteristic length scale depends on the dynamics of the wiggliness. If the wiggliness is constant, $L$ scales linearly, while if the wiggliness grows in time, $L$ exhibits slower growth specified by the energy loss parameters. These results agree with linear scaling observed in Minkowski space simulations \cite{martins2006fractal, sakellariadou1990cosmic}. Moreover, the inclusion of a renormalization scale led to a possible regime where small-scale structure grows in time, but is compensated by a decreasing velocity, with the network's total energy density remaining constant and the domain of the scale remaining unrestricted due to energy conservation. Lastly, when the network is subject to both mechanisms, we determined that the averaging scale becomes bounded from above, as a consequence of the network losing energy. 

In power-law expanding universes, we determined that the three scaling regimes were possible, primarily depending on the expansion rate. For slow and intermediate expansion rates one has growing wiggliness solutions, while in fast expansion regimes one has the Nambu-Goto regime. For a specific value of expansion rate, one finds  a full scaling solution. In the absence of energy losses, this expansion rate coincides with the matter era one, whereas the inclusion of energy loss mechanisms decreases this value, possibly reaching the radiation era for an appropriate value of the energy losses. Our results also demonstrate that scaling is more easily achieved in the matter era rather than the radiation epoch, in agreement with numerical simulations  \cite{martins2006fractal} \cite{bennett1990high} \cite{allen1990cosmic}. It is also interesting to note that the three scaling behaviors here determined have also been found for the case of chiral superconducting strings \cite{oliv} and other current-carrying strings \cite{CVOS}. The case where all three mechanisms act on the network was not included in this work for reasons of space. Lastly, one aspect that remains unclear and should be subject to future studies, is the role of the coarse-graining scale, and in particular, the apparent scale dependence of the total energy of the network. 

These analytic results pave the way for a future calibration of the wiggly model. While our results show qualitative agreement with previous Nambu-Goto numerical simulations, the current available data does not allow for a detailed comparison. Our work therefore motivates additional, higher resolution simulations that cover a broad range of expansion rates and that are able to extract accurate numerical diagnostics for the wiggliness.

\section*{Acknowledgements}

This work was financed by FEDER—Fundo Europeu de Desenvolvimento Regional funds through the COM-PETE 2020—Operational Programme for Competitiveness and Internationalisation (POCI), and by Portuguese funds through FCT - Fundação para a Ciência e a Tecnologia in the framework of the project POCI-01-0145-FEDER-028987 and PTDC/FIS-AST/28987/2017. 

\eject

\bibliographystyle{ws-procs961x669}
\bibliography{mg16almeida}

\end{document}